\documentclass[11pt,letterpaper]{article}

\usepackage[a4paper, total={16cm, 24cm}]{geometry}
\usepackage{graphicx}
\usepackage{url}
\usepackage{color}
\begin{document}

\title{Open-ended innovation arm-race in zero-sum games}

\author{Matteo Marsili\\
\small{The Abdus Salam International Centre for Theoretical Physics}\\
\small{Strada Costiera 11, 34151 Trieste, Italy}\\
Leandro Chaves R\^ego\\
\small{Statistics and Applied Math Departament}\\
\small{Universidade Federal do Cear\'a}\\
\small{Building 910, Pici Campus, 60440-900, Fortaleza/CE, Brazil}\\
\small{Management Engineering Graduate Program}\\
\small{Universidade Federal de Pernambuco}\\
\small{Av. da Arquitetura, s/n, Cidade Universit\'aria, 50740-550, Recife/PE, Brazil}}

\date{}
\maketitle

\begin{abstract}
We discuss zero-sum games with open-ended innovation, whereby each player may introduce new strategies. The innovation process is modelled as a draw of new strategies from a distribution. 
It is argued that, when the cost of innovation is vanishingly small, this setting can lead to an everlasting innovation arm-race. In particular, the introduction of new technologies of the more advanced player makes it more profitable for the backward player to innovate in its turn, whereas innovation of the backward player discourages the more advanced one to innovate.
\end{abstract}

Researchers at DeepMind trained AlphaGo on games played by the masters of the ancient game of Go. 
Within roughly 18 months, the algorithm defeated the European Champion Fan Hui in October 2015, and in March 2016, AlphaGo defeated 18-time World Champion Lee Sedol. In this period, AlphaGo learned all the strategies of the masters and then it started developing new ones, that were never seen in more than 25 centuries. AlphaGo developed even more sophisticated strategies, achieving superhuman performance, by playing millions of games against copies of itself through deep reinforcement learning~\cite{silver2017mastering}. Such spectacular open-ended innovation process would have been impossible for humans. 

Likewise, technological innovation in warfare has witnessed a surge powered by advances in drone technology, Artificial Intelligence and Data Science~\cite{borchert2024very}. The ongoing conflict in Ukraine has accelerated this process further, reorienting a large part of Ukraine's industrial and educational ecosystems~(see V. Goncharuk in \cite{borchert2024very}). Such progresses triggered analogous changes in the Russian military industry, leading to a technological arm-race. 

Similar arm-races powered by AI are going on in other sectors, like the legal industry~\cite{national_law_review_2025_legal_arms_race} and cybersecurity~\cite{oesch2025agentic}, in which competition has made it unacceptable not to adopt AI tools.

Conceptually, these phenomena can be described as zero-sum (or constant-sum) games played repeatedly by two players. 

Our aim is to analyse an open-ended innovation process in zero-sum games, where each player can innovate and introduce new strategies. We show that when the cost of innovation is very low, this can lead to an endless process with vanishingly small payoffs. 

While R\&D competition has been extensively studied in economic literature, traditional models focus primarily on market settings where many firms compete for market share or consumer surplus in non-zero-sum environments~\cite{loury1979market,dasgupta1980industrial,aghion2001competition}. In those settings, innovation can expand aggregate value. Here, we consider a fundamentally different setup: an open-ended innovation race embedded in a two-player zero-sum game. In other words, we focus on pure adversarial conflict -- such as military technological arms races, AI-driven cybersecurity, or zero-sum litigation -- where any strategic gain by one party represents an exact loss for the opponent. We describe such a setting drawing on the theory of large random games~\cite{berg1998matrix} within a comparative statics perspective. 

Remarkably, despite structural differences, we find results that bear striking parallels to key phenomena in industrial organization. For example, we find that technological advances by the leader increase the backward player's marginal utility to innovate, similar to the ``catch-up'' phenomena observed in step-by-step market R\&D races~\cite{aghion2001competition}. 
Furthermore, under vanishing marginal costs of innovation we find an endless expansion of arsenals that mirrors the outcome of classical conflict theory~\cite{hirshleifer1991technology} with highly sensitive conflict technologies. Yet while this outcome in the latter arises from the macroeconomic trade-off between productive effort and appropriative conflict~\cite{hirshleifer1991technology}, in our case it arises purely from the structural incentives of the zero-sum game setting, that trap players in an endless innovation arm race.


\section{Results}

A zero-sum game~\cite{vonneumann1944} is a contest played by two parties where the gain of one of them coincides with the loss of the other. Formally, it is described by a function $\pi(s_1,s_2)$ that quantifies the payoff of player one, when it\footnote{As we refer to possibly artificial players, or to contexts where players are objectified, the neutral gender seems more appropriate.} plays strategy $s_1$ against the opponent strategy $s_2$. Strategies $s_i$ are probability distributions over a simplex 
\begin{equation}
\mathcal{S}_i=\left\{s_i=(s_i^{(1)},\ldots,s_i^{(n_i)}),~s_i^{(j)}\ge 0:~\sum_{j=1}^{n_i} s_i^{(j)} =1\right\},~~~i=1,2
\end{equation}
of dimension $n_i$. In simple terms, player $i$ has $n_i$ possible actions and $s_i^{(j)}$ is the probability with which action $j$ is played in strategy $s_i$. \textcolor{black}{In what follows, the player with more strategies (higher $n_i$ value) is called the advanced player and the other one the backward player.} The payoff of player one is the expected value over these strategies of the payoff $a_{i,j}$ that player $1$ would receive by playing the $i^{\rm th}$ action when player $2$ responds by its $j^{\rm th}$ action, i.e. 
\begin{equation}
\pi(s_1,s_2)=\sum_{i=1}^{n_1}\sum_{j=1}^{n_2}s_1^{(i)} a_{i,j} s_2^{(j)}
\end{equation}
By definition, the payoff of player $2$ -- the opponent -- when it plays strategy $s_2$ against strategy $s_1$, is given by $-\pi(s_1,s_2)$, so that the sum of the payoffs of players is zero. For zero-sum games, 
game theory predicts~\cite{vonneumann1944} that \textcolor{black}{players will follow the maximin paradigm of choosing the strategy that gives the highest payoff assuming that the opponent always uses the strategy most unfavourable to them. In this outcome,} 
the payoff of player $1$ is given by the maximin value~\cite{vonneumann1944}
\begin{equation}
\label{eqV}
V=\pi(s_1^*,s_2^*)=\max_{s_1\in\mathcal{S}_1}\min_{s_2\in\mathcal{S}_2}\pi(s_1,s_2)\,.
\end{equation}
\textcolor{black}{This result assumes that players are rational, but we do not need to make strong assumptions on the rationality of players (as e.g. common knowledge of rationality). All we need to assume is that players defend themselves from the worst that can happen to them without even knowing the opponent's payoffs. Also, we don't need to assume that the solution $(s_1^*,s_2^*)$ is unique, because the payoff of player $1$ is given by the maximin value of Eq.~(\ref{eqV}) in all solutions.
Indeed, Eq.~(\ref{eqV}) also holds for boundedly rational players, because it gives the average payoff obtained by player $1$ under replicator dynamics of evolutionary game theory~\cite{hofbauer2003evolutionary}, which describes also boundedly rational players.}

We consider zero-sum games among players who have a large number of strategies, as appropriate for the contexts discussed above (AI players or industrialised countries). This corresponds to the limit $n_1,n_2\to\infty$ with $n_1/n_2=x$ finite. Innovation can be modelled as the expansion of the set of strategies by the players ($n_i\to n_i+1$), which corresponds to the addition of a column or of a row to the matrix $\hat A$ with entries $a_{i,j}$. In order to capture a genuine open-ended process of innovation, we assume that the entries $a_{i,j}$ of the matrix $\hat A$ are drawn randomly from a distribution. Notice that, although new strategies are drawn at random independently, those which are actually used in equilibrium strategies $s_i^*$ are not independent from those which are already adopted.

\begin{figure}
\begin{center}\includegraphics[width=0.55\linewidth]{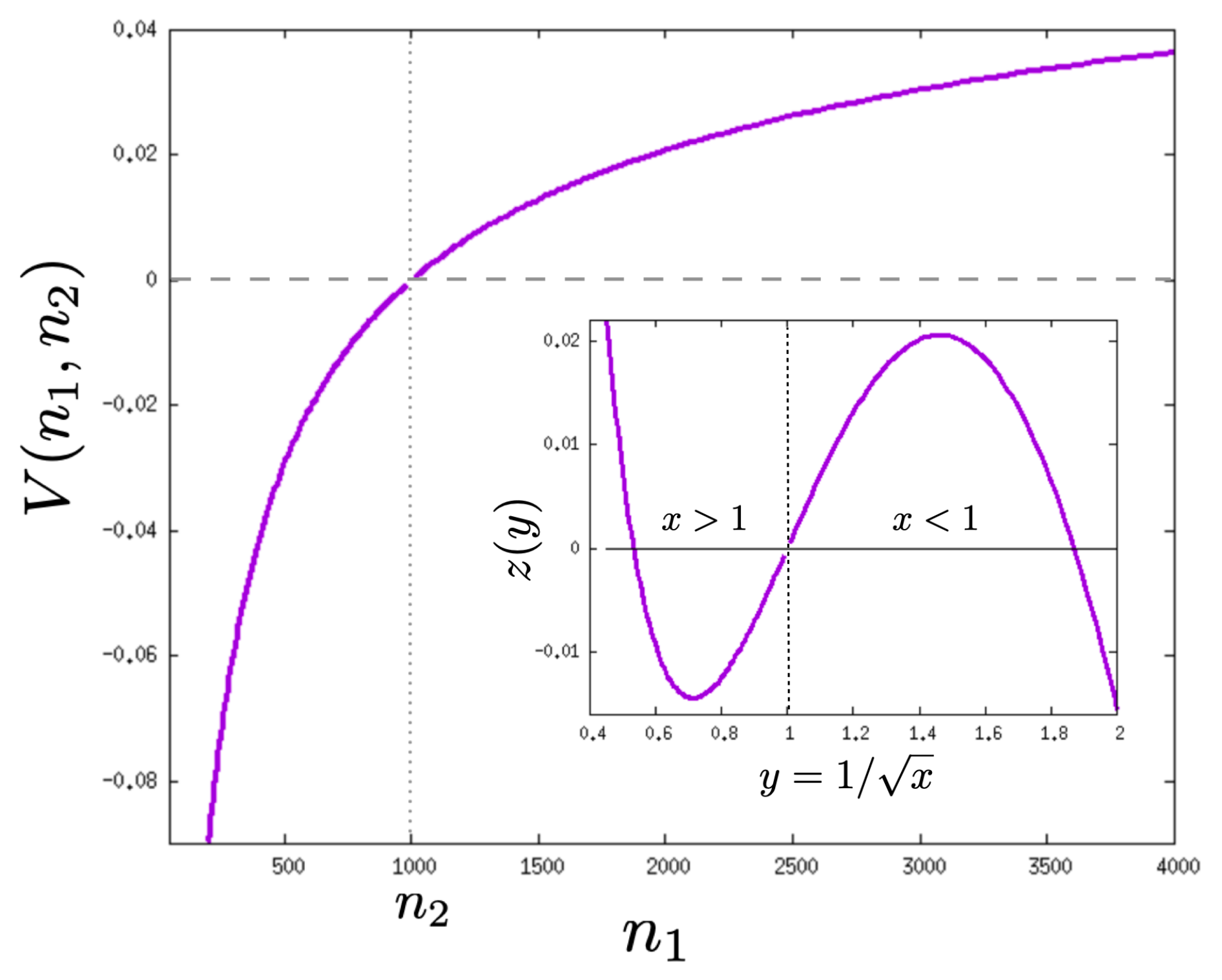}\end{center}
\caption{Value of the game as a function of $n_1$ for $n_2=10^3$ in the BE solution. Inset: plot of 
$z(y)=u(1/y^2)+b(y-1)$ versus $y=1/\sqrt{x}$, with $b$ a constant (adjusted in order to make the convexity more evident). \textcolor{black}{The second derivative $z"(y)$ is proportional to $\frac{3}{2} u'(x)+xu"(x)$, hence the sign of the term in square brackets of Eq.~(\ref{d2V}) can be evinced from the concavity of $z(y)$.}
Both plots are obtained solving the equations in BE for random zero-sum games where $a_{i,j}$ is drawn i.i.d. from a distribution with mean zero and unit variance.}
\label{fig1}
\end{figure}

Hence, we consider large random zero-sum games, as discussed by Berg and \textcolor{black}{Engel} (BE)~\cite{berg1998matrix}, where $a_{i,j}$ is the $i,j$ element of a $n_1\times n_2$ random matrix $\hat A$, which is drawn from a distribution with zero mean and finite variance. The results of BE~\cite{berg1998matrix} show that, in the limiting case where $n_1,n_2\to\infty$ and $n_1/n_2= x$,
the value of the game in Eq. (\ref{eqV}) is a self-averaging quantity, which is well approximated by\footnote{BE consider the asymptotic regime in which the dimensions of the payoff matrix grow proportionally, i.e., $n_1,n_2\to\infty$ with $n_2/n_1\to\alpha$. Under the assumptions of a random payoff matrix adopted in their analysis, the normalized value of the game converges to a deterministic quantity that depends on the matrix dimensions only through the aspect ratio ($\alpha$):
\[
\nu_c(\alpha)=\max_{x_1,\ldots,x_{n_1}}\min_{j\in[1,n_2]}\sum_{i=1}^{n_1}x_i c_{i,j}\,.
\]
In the present notation, $x_i=n_1 s_1^{(i)}$, $c_{i,j}=a_{i,j}/\sqrt{n_1}$ and $\alpha=n_2/n_1=1/x$. Therefore $\nu_c(\alpha)=\sqrt{n_1}V(n_1,n_2)$ which implies that the function $u$ is related to $\nu_c$ in BE by $u(x)=\nu_c(1/x)/\sqrt{x}$, or $u(x)=-\nu_c(x)$ by Eq.(\ref{equx}).} 
\begin{equation}
\label{eqVN}
V(n_1,n_2)=\frac{1}{\sqrt{n_2}}u\left(\frac{n_1}{n_2}\right)\,.
\end{equation}
The precise form of the function $u(x)$ depends on the specific distribution from which the matrix elements are assumed to be drawn. Yet this function \textcolor{black}{satisfies} few general properties that are sufficient for our goals: first, $u(1)=0$, because this is a case where both players are equipped with an equivalent arsenal of strategies\textcolor{black}{, and hence $V(n_1,n_2)=-V(n_1,n_2)=0$}. Second, \textcolor{black}{interchanging the roles of the two players, we find that $V(n_1,n_2)$ satisfies the symmetry relation} $V(n_1,n_2)=-V(n_2,n_1)$. This implies the relation
\begin{equation}
\label{equx}
u(x)=-\frac{1}{\sqrt{x}}u\left(1/x\right)\,.
\end{equation}
Furthermore, $V(n_1,n_2)$ decreases with $n_2$ and increases with $n_1$. Hence $u(x)$ is an increasing function of $x$. Lastly, $V(n_1,n_2)$ is a concave function of 
$n_1$, i.e. $\frac{\partial^2 V}{\partial n_1^2}<0$. This is definitely true in a neighbourhood of $n_1=n_2$, because $u"(1)=-\frac 3 2 u'(1)<0$. BE's analysis shows that $u(x)$ is a concave function for all values of $x$, as shown in Fig.~\ref{fig1}. For $n_1\neq n_2$, concavity amounts to assuming that the payoff increase that player one gains from innovation decreases with $n_1$, which is consistent with decreasing marginal returns to R\&D investment~\cite{Bloom2020}\textcolor{black}{. In simple terms, decreasing marginal returns to innovation means that each additional innovation yields smaller gains -- so the more a player innovates, the less profitable it becomes to innovate further}. 

\textcolor{black}{Instead,} innovation of the most advanced player increases the \textcolor{black}{incentives of the backward player to innovate
.} In order to see this, imagine that player $2$ innovates, i.e. $n_2\to n_2+\delta n_2$. 
\textcolor{black}{We focus on the effect on the marginal gain of player $1$ to innovate, which is the increase or decrease in $\frac{\partial V}{\partial n_1}$. If this increases (resp., decreases), this means that innovation by player 2 increases (resp., decreases) the incentives of player one to innovate.}
The change in the marginal gain of player $1$ to innovate is given by
\begin{equation}
\label{d2V}
\delta\frac{\partial V}{\partial n_1}=\frac{\partial^2 V}{\partial n_1\partial n_2}\delta n_2=-\left[\frac{3}{2} u'(x)+xu"(x)\right]\frac{\delta n_2}{n_2^{5/2}}\,,\qquad x=\frac{n_1}{n_2}\,.
\end{equation}
\textcolor{black}{The term in brackets vanishes if $x=0$, is positive if $x>1$ and is negative if $x<1$. This is shown in the Appendix and, graphically in the inset of Fig.~\ref{fig1} for the BE solution (see caption). This means that, if player 1 is the advanced player ($n_1>n_2$, i.e. $x>1$), then an innovation $\delta n_2$ of player 2 (the backward one) decreases its marginal gain $\delta\frac{\partial V}{\partial n_1}<0$. Conversely, if player 1 is the backward one ($x<1$), innovation of the advanced player provides more incentives for it to innovate.} \textcolor{black}{Thus, the reward of innovating increases as the opponent innovates only if the opponent is already in a better strategic position of having more strategies. Otherwise, innovation of an opponent who is in a worse strategic position makes innovation less compelling for the more advanced player.}

\textcolor{black}{Let us assume that the cost of innovation is proportional to $\delta n_i$. Then player 1 will innovate} as long as the gain $V(n_1+\delta n_1,n_2)-V(n_1,n_2)\simeq \frac{\partial V}{\partial n_1}\delta n_1$ exceeds the cost $c_1\delta n_1$, \textcolor{black}{where $c_1$ is the marginal cost of innovation. If the same holds for player $2$, with marginal cost $c_2$, then an innovation}
arm-race will continue for as long as $n_i\to n_i^*$ with
\begin{equation}
\label{eqnistar}
\frac{\partial V}{\partial n_1}=(n_2^*)^{-3/2}
u'\left(n_1^*/n_{2}^*\right)=c_1\,~~~\hbox{and}~~~
\frac{\partial (-V)}{\partial n_2}=(n_1^*)^{-3/2}u'\left(n_2^*/n_{1}^*\right)=c_2\,.
\end{equation}
\textcolor{black}{Note that, because of the convexity of $u(x)$, if} the costs of innovation \textcolor{black}{vanish} ($c_i\to 0$), the asymptotic number of strategies diverges $n_i^*\to\infty$. \textcolor{black}{We finally show that in the limit when both $c_1,c_2\to 0$, with $c_1/c_2$ finite}, the game will never stop, open-ended innovation will proceed endlessly.

\textcolor{black}{In order to see this, note that} the ratio between the two Equations~(\ref{eqnistar}) determines 
$x^*=n_1^*/n_2^*$, i.e.
\begin{equation}
\label{neweqt}
\textcolor{black}{\frac{c_1}{c_2}=(x^*)^{3/2}\frac{u'(x^*)}{u'(1/x^*)}\,.}
\end{equation}
\textcolor{black}{Differentiating Eq. (\ref{equx}), it follows that:}
\begin{equation}
\label{neweqt2}
\textcolor{black}{u'(x^*)=\frac{1}{2}(x^*)^{-3/2}u(1/x^*)+(x^*)^{-5/2}u'(1/x^*)\,.}
\end{equation}
\textcolor{black}{Isolating $u'(1/x^*)$ in Eq. (\ref{neweqt2}) and replacing it in Eq. (\ref{neweqt}), it follows that:}
\begin{equation}
\frac{c_1}{c_2}=\frac{u'(x^*)}{x^* u'(x^*)+u(x^*)/2}\,.
\end{equation}
It can be shown that $x^*=n_1^*/n_2^*$ is an increasing function of $c_2/c_1$. The player with the lowest marginal cost will be the one with a higher repertoire of strategies, and the one that will gain a positive payoff in the end. Note that, if $c_1/c_2$ is finite then $x^*$ is also finite. In the limit $c_1, c_2\to 0$ with $c_1/c_2$ finite, Eq.~(\ref{eqnistar}) implies that $n_1^*,n_2^*\to\infty$, i.e. the innovation arm-race will never stop.

\section{Conclusion}

Repeated zero-sum games of players who can innovate introducing new strategies can lead to endless open-ended innovation when the cost of innovation is vanishingly small. This condition applies in particular when the innovation process can be automated, by the use of AI, or when players are countries subject to existential threats claiming all of its economy's resources for defense. The arguments presented here suggest that, in these conditions, the end result of such prolonged open-ended innovation arm race is vanishingly small payoffs for both players. 

\textcolor{black}{We hope that the results of this paper provide formal arguments to support regulatory approaches in contexts where AI driven unbridled innovation can spiral out of control at the detriment of human life.}



\paragraph*{Acknowledgments}
Discussions with Johannes Berg are gratefully acknowledged.

\appendix

\section{Opponent's innovation decreases marginal payoff}

The marginal payoff of player $1$ is 
\[
\frac{\partial V}{\partial n_1}=n_2^{-3/2}u'(n_1/n_2)\,.
\]
It's change, when $n_2\to n_2+\delta n_2$, is given by Eq.~(\ref{d2V}). The function $u(x)$ can be expressed in terms of a function of $t=\log x$ defined as $g(t)=x^{1/4}u(x)$ which, by virtue of Eq.~(\ref{equx}) is anti-symmetric, i.e. $g(t)=-g(-t)$, as well as all its even derivatives. In terms of this function we have 
\begin{equation}
    \frac{\partial^2 V}{\partial n_1\partial n_2}=\left[\frac{1}{16}g(t)-g"(t)\right](n_1 n_2)^{-5/4}=-\left[\frac 3 2 u'(x)+xu"(x)\right]n_2^{-5/2}\,.
\end{equation}
The first equality implies that, upon exchanging players (i.e. $n_1\leftrightarrow n_2$ or $t\leftrightarrow -t$), $\frac{\partial^2 V}{\partial n_1\partial n_2}$ changes sign and it vanishes for $n_1=n_2$, because all even derivatives of $g(t)$ are odd under the change $t\leftrightarrow -t$. 

As for the second equality, given that $\frac 3 2 u'(x)+x u"(x)=\frac{y^4}{4}z"(y)$ where $z(y)=u(x)+b(y-1)$ with $y=1/\sqrt{x}$ and $b$ a constant, the sign of $\frac{\partial^2 V}{\partial n_1\partial n_2}$ can be read from the concavity of the function $z(y)$, as shown in the inset of Fig.~\ref{fig1}.

\bibliographystyle{plain} 
\bibliography{war_open_ended_innovation} 

\end{document}